\documentclass{elsarticle}
\usepackage[utf8]{inputenc}

\usepackage{adjustbox}
\usepackage[toc,acronym]{glossaries}

\makeglossaries

\newacronym{AI}{AI}{Artificial Intelligence}
\newacronym{ALU}{ALU}{Arithmetical and Logical Unit}
\newacronym{CPU}{CPU}{Central Processing Unit}
\newacronym{FPGA}{FPGA}{Field Programmable Gate Array}
\newacronym{FIFO}{FIFO}{First In/First Out storage}
\newacronym{HW}{HW}{hardware}
\newacronym{ISA}{ISA}{Instruction Set Architecture}
\newacronym{I/O}{I/O}{Input/Output}
\newacronym{HPC}{HPC}{Hight Performance Computing}
\newacronym{ICCB}{ICCB}{Inter-Core Communication Block}
\newacronym{LAN}{LAN}{Local Area Network}
\newacronym{MC}{MC}{Multi-Core and/or Many-Core}
\newacronym{MLP}{MLP}{Memory Level Parallelism}
\newacronym{OoO}{OoO}{Out-of-Order}
\newacronym{OS}{OS}{operating system}
\newacronym{PD}{PD}{Propagation Delay}
\newacronym{QT}{QT}{Quasi-Thread}
\newacronym{PU}{PU}{Processing Unit}
\newacronym{SPA}{SPA}{Single Processor Approach}
\newacronym{SW}{SW}{software}
\newacronym{HPL}{HPL}{High Performance Linpack}
\newacronym{HPCG}{HPCG}{High Performance Conjugate Gradients}

\newacronym{EMPA}{EMPA}{Explicitly Many-Processor Approach}
\newacronym{EPE}{EPE}{EMPA Processing Element}
\newacronym{EME}{EME}{EMPA Morphing Element}
\newacronym{ECE}{ECE}{EMPA Communicating Element}
\newacronym{EICB}{EICB}{EMPA Inter-Core Block}
\newacronym{ESME}{ESME}{EMPA Storage Manager Element}
\usepackage {xcolor}

\definecolor{webgreen}{rgb}{0,.5,0}
\definecolor{webbrown}{rgb}{.6,0,0}
\definecolor{webyellow}{rgb}{0.98,0.92,0.73}
\definecolor{webgray}{rgb}{.753,.753,.753}
\definecolor{webblue}{rgb}{0,0,.8}
\definecolor{webgreen}{rgb}{0, 0.5, 0} % less intense green
\definecolor{webred}{rgb}{0.5, 0, 0}   % less intense red

\usepackage{hyperref}

\journal{Neurocomputing}

\begin{document}

\begin{frontmatter}

\title{Do we know the operating principles of our computers\\
	 better than those of our brain?}

\author%[mymainaddress]
{J\'anos V\'egh\corref{mycorrespondingauthor}}
\address{Kalim\'anos BT, Hungary}
\ead{Vegh.Janos@gmail.com}

\author%[mysecondaryaddress]
{\'Ad\'am J. Berki}
%\ead{berki.adam@yahoo.com}
\address[mysecondaryaddress]{%George Emil Palade 
	University of Medicine, Pharmacy, Sciences and Technology of Targu Mures, Romania}

%\address[mymainaddress]{Kalim\'anos BT, Hungary}
\cortext[mycorrespondingauthor]{Corresponding author}

\begin{abstract}
 The increasing interest in understanding the behavior of the biological neural
networks, and the increasing utilization of artificial neural networks in different
fields and scales, both require a thorough understanding of how neuromorphic
computing works. On the one side, the need to program those artificial neuron-like
elements, and, on the other side, the necessity for a large number of such elements to cooperate, communicate and compute during tasks, need to be scrutinized to determine how
efficiently conventional computing can assist in implementing such systems.
Some electronic components bear a surprising resemblance to some biological
structures. However, combining them with components that work using different
principles can result in systems with very poor efficacy. The paper discusses how the conventional principles, components and thinking about computing limit mimicking the biological systems. We describe what changes will be necessary in the computing paradigms to get closer to the
marvelously efficient operation of biological neural networks.		
\end{abstract}


\end{frontmatter}

%\linenumbers

\tableofcontents

\section{Introduction}

Today we have the "golden age" of neuromorphic (brain-inspired, artificial intelligence)  architectures and computing. However, the meaning of the word has changed considerably since Carver Mead~\cite{NeuromorphicSystems:1990}  coined the wording. Today practically every single solution that borrows at least one single operating principle from the biology and mimics some of its functionality in a more or less successful way deserves this name. As always, to grasp out some single aspect and implement it in an environment and from components based on entirely different principles, is dangerous. Historically, 'neuromorphic' architectures were suggested to be based on different principles and components, such as mechanics, pneumatics, telephones, analog and digital electronics, computing. Some initial resemblance surely exists, and even some straightforward systems can demonstrate more or less successfully functionality in some aspects similar to that of the nervous system. There is a noteworthy analogy between the deep learning of neuronal nodes and the long-term potentiation found in synapses.

However, when scrutinizing the scalability (i.e., how those systems shall work when used under real-life conditions in which a vast number of similar subsystems shall work and cooperate), the picture is not favorable at all.
"\textit{Successfully addressing these challenges [of neuromorphic computing] will lead to a new class of computers and systems architectures}"~\cite{NeuromorphicComputing:2015} has been targeted. However, as noticed by the judges of the Gordon Bell Prize, "\textit{surprisingly, [among the winners,] there have been no brain-inspired massively parallel specialized computers}"~\cite{GordonBellPrize:2017}. Despite the vast need and investments, furthermore the concentrated and coordinated efforts, just because of mimicking the biological systems with computing inadequately.

Given "\textit{that the quest to build an electronic computer based on the operational principles of biological brains has attracted attention over many years}"~\cite{FurberNeuralEngineering:2007}, modeling the neuronal operation became
a well-known field in electronics. At the same time, more and more details come to light about the 
computational operations of the brain. However, it would appear, that the 'wet' neuroscience
is miles ahead of the 'silicon' neuroscience. 
There are projects and exaggerated claims about extremely large computing systems,
even about targeting the simulation of the
brain of some animals and eventually even the human brain. 
Often these claims are followed by a long silence, or some rather slim or no results.
As that the operating principles of the large computer systems tend to deviate from the operating principles of a single processor,
it is worth reopening the discussion on a decade-old question
"\textit{Do computer engineers have something to contribute. . . to the understanding of brain and mind?}"
~\cite{FurberNeuralEngineering:2007}.
Maybe, and they surely have something to contribute to the understanding of computing itself.
\textit{There is no doubt that the brain does computing, the key question is how?}

Section~\ref{sec:LargeScale} presents some %tragically low efficacy
computing systems, having, as we point out: as a consequence of the computing paradigm, enormously high energy consumption.
Section~\ref{sec:SingleProc} discuses the primary reasons for the issues and failures: the computing paradigm and their consequences such as
the serial bus and the effect of the physical size.
The timely behavior is especially important in the biological objects, so their fair imitation in the computing systems is of crucial importance, as section~\ref{sec:timely} discusses it.
The neuromorphic computing is a special type of workloads
that have a dominating role in forming the computational efficiency of the computing systems, as section~\ref{sec:Workload} discusses it.
Section~\ref{sec:ClassicComputing} presents some further limitations,
rooting in the classical paradigm; furthermore, it draws parallels
with classic versus modern science and classic versus modern computing.
Section~\ref{sec:NotSimple} provides examples, why is of limited validity
to consider the role of a grasped-out component: a neuromorphic system is not
a simple sum of its components.
In section~\ref{sec:Summary}, the paper attempts to make a clear pointer
where we can continue using the classic computing and where we shall base the systems on the new principles.

\section{Issues with the large scale computing\label{sec:LargeScale}}

 The worst limiting factor in conventional computing is the method of communication between processors, which increases exponentially with increasing complexity/number. Historically in the model of computing proposed by von Neumann, there is one single entity, an isolated (non-communicating) processor, whereas in the bio-inspired models, billions of entities, organized into specific assemblies, cooperate via communication. (The communication here means not only sending data, but also sending/receiving signals, including synchronization of the operation of the entities.)
 Neuromorphic systems, expected to perform tasks in one paradigm, but assembled from components manufactured using principles of (and implemented by experts trained in) the other paradigm are unable to perform at the required speed and efficacy for real-world solutions. The larger the system, the higher the communication load and the performance debt.

To get nearer to the marvelously efficient operation of the biological brain,  other features must also be mimicked from the biology. Only a little portion of the neurons are working simultaneously in solving the actual task; there is a massive number of very simple ('very thin') processors rather than a 'fat' processor\footnote{One should scrutinize whether it is worth to implement accelerators (such as pipelines, branch predictors) intended to be used in large computing systems, to achieve just a couple of times higher processing speed at the price of using several hundred times more transistors}; only a portion of the functionality and connection are pre-wired, the rest is mobile;
there is an inherent redundancy, replacing a faulty neuron may be possible via systematic training. The conventional
processors can only either run or halt, but not to make a little break.
The biology uses purely event-driven (asynchronous) computing, while modern electronics uses clock-driven systems; for the catastrophic consequences of attempting to simulate a neuromorphic system, such as the human brain, using components prepared
for conventional computing,
see the case of SpiNNaker, discussed below.

The large computing systems can cope with the tasks with growing difficulty, enormously decreasing computing efficiency, and enormously growing energy consumption.
Being not aware of that the collaboration between processors
needs a different approach (another paradigm),
resulted
in demonstrative failures already known (such as the supercomputers Gyoukou and Aurora'18, or the brain simulator SpiNNaker)\footnote{The explanations are quite different: Gyoukou is simply withdrawn; Aurora is practically
	withdrawn: retargeted and delayed; Despite the failure of SpiNNaker1, the SpiNNaker2 is also under construction~\cite{SpiNNaker2:2018};
	"Chinese decision-makers decided to withhold the country’s newest Shuguang supercomputers even though they operate more than 50 percent faster than the best current US machines".} and many more (all they intend to deliver 0.13-0.2~Eflops) may follow:  such as Aurora'21~\cite{DOEAurora:2017},
the China mystic supercomputers\footnote{https://www.scmp.com/tech/policy/article/3015997/china-has-decided-not-fan-flames-super-computing-rivalry-amid-us} and
the EU planned supercomputers\footnote{https://ec.europa.eu/newsroom/dae/document.cfm? doc\_id =60156}.
Systems having "only" millions of processors already show the issues, and the brain-like systems want to comprise four orders of magnitude higher number of computing elements. Besides, the scaling is strongly nonlinear~\textbf{\cite{VeghReevaluate:2020,VeghScalingANN:2020}}. When targeting neuromorphic features such as "deep learning training", the issues start to manifest at just a couple of dozens of processors~\cite{DeepNeuralNetworkTraining:2016}\textbf{\cite{VeghAIperformance:2020}}.

\section{Limitations due to the Single Processor Approach\label{sec:SingleProc}}

As suspected by many experts, the computing paradigm itself, "\textit{the implicit   hardware/software  contract}~\cite{AsanovicParallelCACM:2009}", is responsible for the experienced issues:
"\textit{No current programming model is able to cope with this development [of processors], though, as they essentially still follow the classical van Neumann model}"~\cite{SoOS:2010}.
When thinking about "advances beyond 2020", the solution was expected from the "\textit{more efficient implementation of the von Neumann architecture}"~\cite{DeBenedictis_supercomputing:2014}, however.
Even when speaking about
building up computing from scratch ("rebooting the model"~\cite{RebootingComputingModels:2019}), only implementing different gating technology for \textit{the same computing model} is meant. However, the paradigm prevents
building large neuromorphic systems, too.

\begin{figure*}
	\maxsizebox{\textwidth}{!}
	{
		\includegraphics[width=\textwidth]{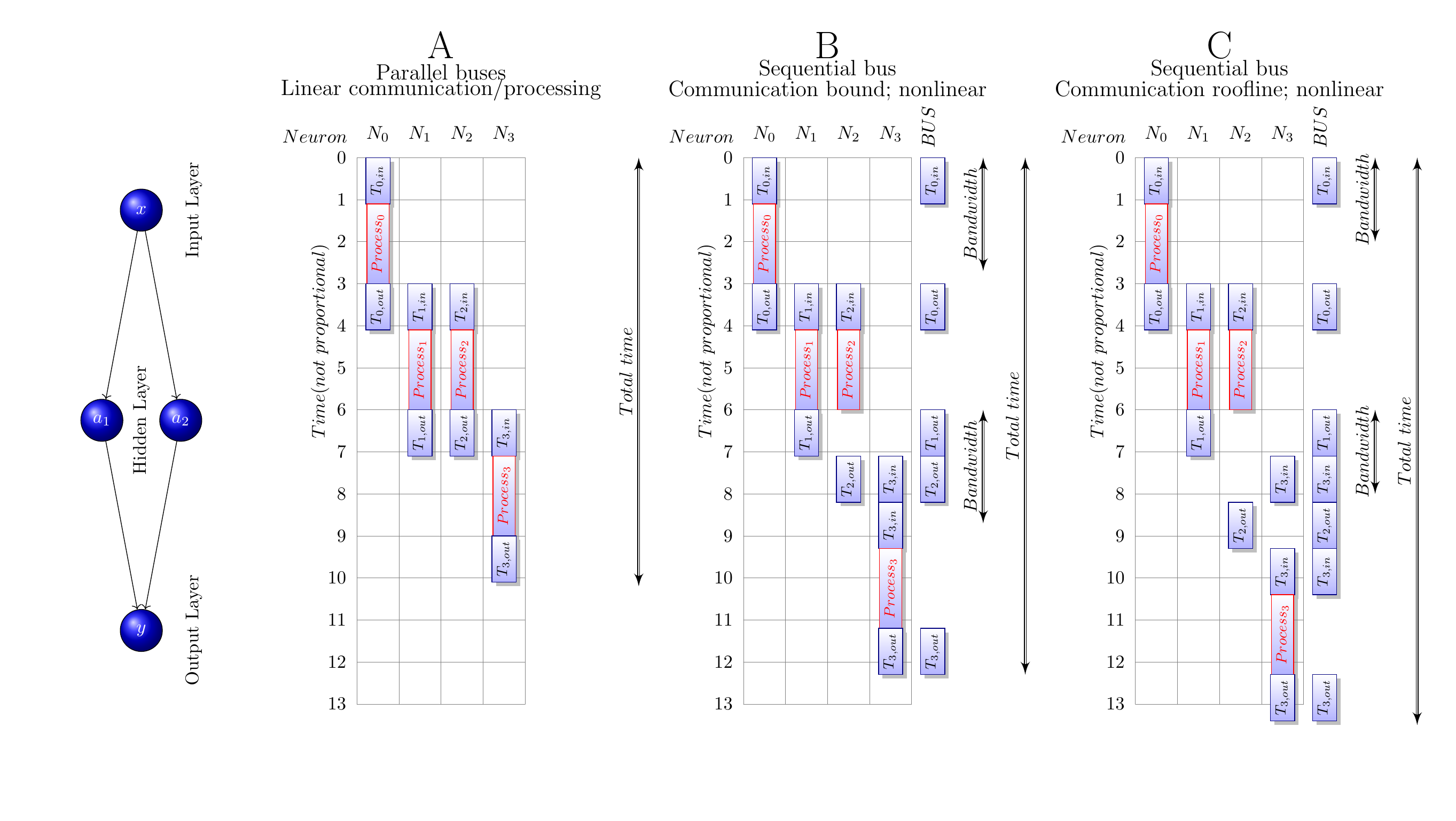}
	}
	\vspace{-\baselineskip}

	\caption{Implementing neuronal communication in different technical approaches. A: the parallel bus; B and C: the shared serial bus, before and after reaching the communication "roofline"~\cite{WilliamsRoofline:2009}\label{fig:Neuronal}
	}
\end{figure*}

 {The bottleneck is essentially the "technical implementation" of the communication, stemming from the \gls{SPA}, as illustrated in Fig.~\ref{fig:Neuronal}.
	The inset shows a simple neuromorphic use case: one input neuron and one output neuron communicating through a hidden layer, comprising only two neurons.
	Fig.~\ref{fig:Neuronal}.A mostly shows the biological implementation: all neurons are directly wired to their partners, i.e.,
	a system of "parallel buses" (the axons) exists. Notice that the operating time also comprises two non-payload times: the data input and data output, which coincide with the non-payload time of the other communication party. The diagram displays the logical and temporal dependencies of the neuronal functionality.

The payload operation ("the computing") can only start after the data is delivered (by the, from this point of view,
	non-payload functionality: input-side communication), and the output communication can only begin when the computing finished. Importantly, the communication and calculation mutually block each other.
	Two important points that neuromorphic systems must mimic noticed immediately: i/ \textit{the communication time is an integral part of the
		total execution time}, and ii/ \textit{the ability to communicate is a native functionality} of the system.
	In such a parallel implementation, \textit{the performance of the system}, measured as the resulting total time (processing + transmitting), \textit{scales linearly with increasing both the non-payload communication speed and the payload processing speed}.
}

	The present technical approaches assume a similar
	linearity of the performance  of the computing systems as ”\textit{Gustafson’s formulation~\cite{Gustafson:1988}
		gives an illusion that as if N [the number of the processors] can increase indefinitely}”~\cite{AmdalVsGustafson96}.
	The fact that ”\textit{in practice,
		for several applications, the fraction of the serial part happens
		to be very, very small thus leading to \textbf{near-linear} speedups}”~\cite{AmdalVsGustafson96}, however, misled the researchers.
	\textit{Gustafson's
		’linear scaling’ neglects the communication entirely} (which is not the case, especially not in neuromorphic computing). He established his conclusions on only several hundred processors, and
	the interplay of the improving parallelization and the general
	\gls{HW} development (including the non-determinism
	of the modern \gls{HW}~\cite{PerformanceCounter2013}) covered for decades that \textit{the scaling
		was used far outside of its range of validity}~\textbf{\cite{VeghReevaluate:2020,VeghScalingANN:2020}}.
	Not considering the effect of the time of communication (i.e., the timely behavior), means not considering a vital feature of the biological system. Essentially the same effect
	(the vastly increased number of idle cycles due to the physical size of supercomputers) leads to the failures of supercomputer projects (for a detailed discussion, see~\textbf{\cite{VeghHowMany:2020}}).
	\textit{The 'real scaling' is strongly nonlinear, with nature-defined bound.}

	Fig.~\ref{fig:Neuronal}.B shows a \textit{technical implementation of  a high-speed shared bus} for communication.
	To the right of the grid, the activity that loads the bus at the given time is shown.
	A double arrow illustrates the communication bandwidth, the length of which is proportional to the number of packages the bus can deliver in a given time unit.
	The high-speed bus is only very slightly loaded.
	We assume that the input neuron can send its information in a single message to the hidden layer furthermore that the processing by the neurons in the hidden layer both starts and ends at the same time. However, the neurons must compete for accessing the bus, and only one of them can send its message immediately, the other(s)
	must wait until the bus gets released.
	The output neuron can only receive the message when the first neuron completes it.
	Furthermore, the output neuron must first acquire the second message from the bus, and the processing can only begin after having both input arguments. \textit{This constraint results in sequential bus delays both during non-payload processing in the hidden layer
		and the payload processing in the output neuron}.
	Adding one more neuron to the layer introduces one more delay,
	which explains why "\textit{shallow networks with many neurons per layer \dots scale worse than deep networks with less neurons}"
	\cite{DeepNeuralNetworkTraining:2016}:
	the system sends them at different times in the different layers
	(and even they may have independent buses between the layers), although the shared bus persists in limiting the communication.

\begin{figure}
		\includegraphics[width=\textwidth]{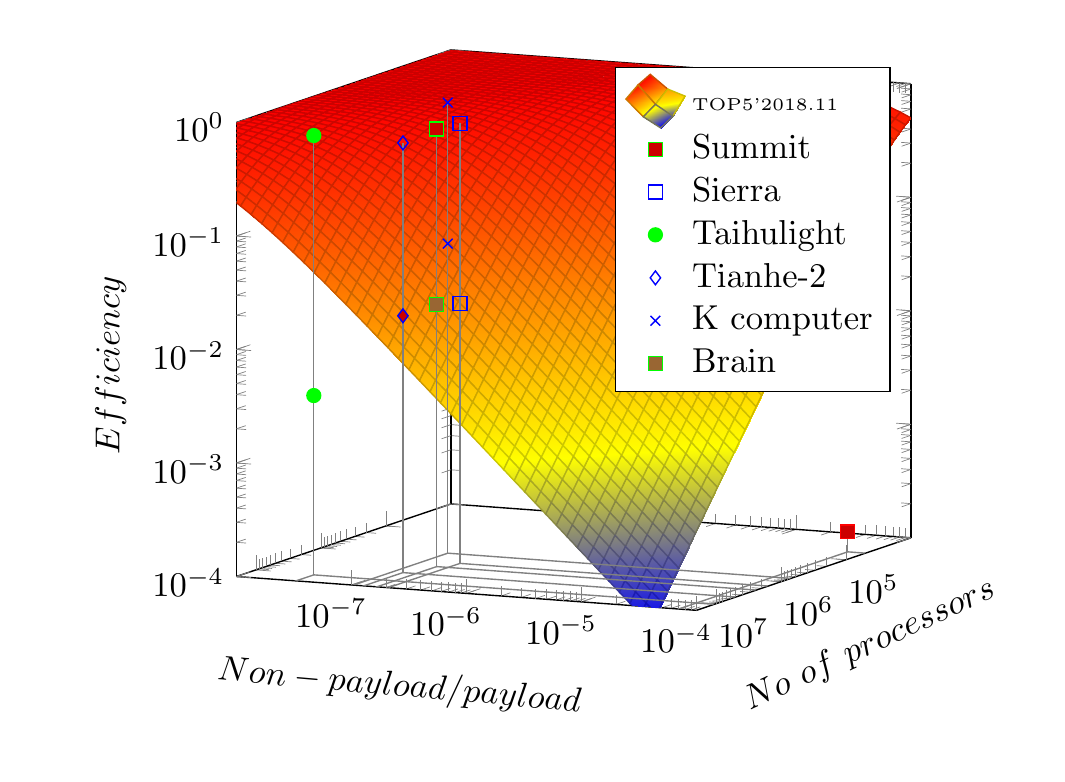}

	\caption{The effect of the \gls{SPA} implementation of the computing performance.
		The surface displays how the efficiency of the \gls{SPA} systems depends on the goodness of the parallelization (aka the non-payload/payload ratio)
		and the number of cores.
		 The figure marks show at what efficiency values can the top supercomputers run different workloads (for a discussion see section~\ref{sec:Workload}): the
		'best load' benchmark \gls{HPL}, and the 'real-life load' \gls{HPCG}.
		The right bottom part displays the expected efficiency of running
		neuromorphic calculations on \gls{SPA} computers.
		\label{fig:EfficiencySPA}
	}
\end{figure}

	The dependence of the performance is strongly nonlinear at higher performance values (implemented using a large number of processors). The effect is especially disadvantageous for the systems,
	such as the neuromorphic ones, that need much more communication, thus making the non-payload to payload ratio very wrong.
	The linear dependence at low nominal performance values explains why the initial successes of \textit{any new technology, material or method} in the field, using the classic computing model, can be misleading: in simple cases, the
	classic paradigm performs tolerably well thanks to that compared to biological neural networks, current neuron/dendrite models are simple,
		the networks small and learning models appear to be rather basic.
	Recall that
	for artificial neuronal networks the saturation is reached
	at just dozens of processors~\cite{DeepNeuralNetworkTraining:2016},
	because of the extreme high proportion of communication.

Fig.~\ref{fig:EfficiencySPA} depicts how
the \gls{SPA} paradigm also defines the computational performance of the
parallelized sequential system. Given that the task defines
how many computations it needs, and the computational time
is inversely proportional with the efficiency,
one can trivially conclude that \textit{the decreasing computational efficiency
leads to increasing energy consumption}, just because of the \gls{SPA}.
 "\textit{This decay in performance is not a fault of the
	architecture, but is dictated by the limited parallelism}"~\cite{ScalingParallel:1993}.

\section{The importance of imitating the timely behavior\label{sec:timely}}
In both biological and electronic systems, both the distance between the
entities of the network, and the signal propagation speed is finite.
Because of this, in the physically large-sized systems the
'idle time' of the processors defines the final performance
a parallelized sequential system can achieve.
In the conventional computing systems also the 'data dependence' limits the available parallelism: we must compute the data before we can use it as an argument for another computation.
Although of course in the conventional computing the data must be delivered
to the place of the second utilization, thanks to the 'weak scaling'~\cite{Gustafson:1988}, this 'communication time' is neglected.

{In neuromorphic computing, however, as discussed in connection with
Fig.~\ref{fig:Neuronal}, the transfer time is a vital part of information processing.
A biological brain must deploy a "speed accelerator" to ensure that the control signals arrive at the target destination
before the arrival of the controlled messages, despite that the former derived from a distant part of the brain~\cite{BuzsakiGammaOscillations:2012}.
\textit{This aspect is so vital in biology that the brain deploys many cells with the associated energy investment 
to keep the communication speed higher for the control signal.}
Computer technology cannot speed up the communication selectively, as in biology, and it is not worth to keep part of the system for a lower speed selectively. 

%The \gls{EMPA} approach, suggested earlier~\cite{VeghSPAEMPA:2020},
%mimics the biology (see Fig.~\ref{fig:AxonBus}B). It introduces directly-wired connections between physically neighboring cells,
%creates a special hierarchical bus system and places a special communication unit, the
%\gls{ICCB}, between the computer cores, mimicking neurons.
%The figure shows the "cluster head" in different color: it has a distinguished role
%as it has access to the far and local memories (M) and can forward messages
%via the (G) gateway to other clusters (similar gateways can be implemented 
%for the interprocessor communication, and higher organizational levels; 
%providing access to different levels of the hierarchical buses).

Extending on our previous work [22], here we introduce the Explicitly Many-Processor Approach (EMPA) which is a new method using clusters of computational units (‘neurons’ Fig.~\ref{fig:AxonBus}B) to mimic the timely behaviour of ‘biological brains’. The clusters are built using the following key novel ideas: 1) implementing directly-wired connections between physically neighbouring cells; 2) creating a special hierarchical bus system; 3) placing a special communication unit, the (\gls{ICCB}, Fig.~\ref{fig:AxonBus}B, purple) between the computer cores mimicking neurons ( Fig.~\ref{fig:AxonBus}B, green); 4) creating a specialized ‘cluster head’ ( Fig.~\ref{fig:AxonBus}B) with the extra abilities to access the local and far memories ( Fig.~\ref{fig:AxonBus}~M) and to forward messages via the gateway ( Fig.~\ref{fig:AxonBus}~ G) to other ‘clusters’ (similar gateways can be implemented for the inter-processor communication, and higher organizational levels; providing access to different levels of the hierarchical buses). 
The cluster members are denoted by their relative position 
(the addressing mode enables using virtual cores, mapped to physical cores at runtime), and they can access the memory and other clusters only through the head of the cluster. This enables both easy sharing of locally important 
state variables, keeps local traffic away from the bus(es) and reduces wiring
inside the chip. 
The \gls{ICCB}s can forward messages via direct wired connections with up to 2 'hops' to the 
immediate neighbors and the second neighbors (even if they belong to another cluster). This solution enables billions of 'neurons' to communicate 
at the same time, without delay, although the distant neuron must use 
one (or more) of the hierarchical buses, in function or their location.
The resemblance between Fig.~\ref{fig:AxonBus}A and Fig.~7 in reference~\cite{BuzsakiGammaOscillations:2012}
underlines the importance of making a clear distinction between handling 'near' and 'far' signals,
and accounting for relative signal timing. Although the \gls{ICCB} blocks can adequately represent 'locally connected' interneurons and the 'G' gateway
the 'long-range interneurons'~\cite{BuzsakiGammaOscillations:2012}, in biological systems conduction time must be separately maintained by the neurons
in biology-mimicking computing systems.
Making time-stamps and relying on the computer network delivery principles is not sufficient
for maintaining correct relative timing. \textit{The timely behavior is a vital feature of the biology-mimicking systems, cannot not be replaced it with the synchronization principles of computing}.
Ignoring this requirement massively contributes to the failures
of biology-mimicking computing systems. Communication time  is less vital
for using neurons in \gls{AI}, but even in that case, one must consider the communication time explicitly.

\begin{figure*}
	\begin{tabular}{cc}
		\includegraphics[width=.50\textwidth]{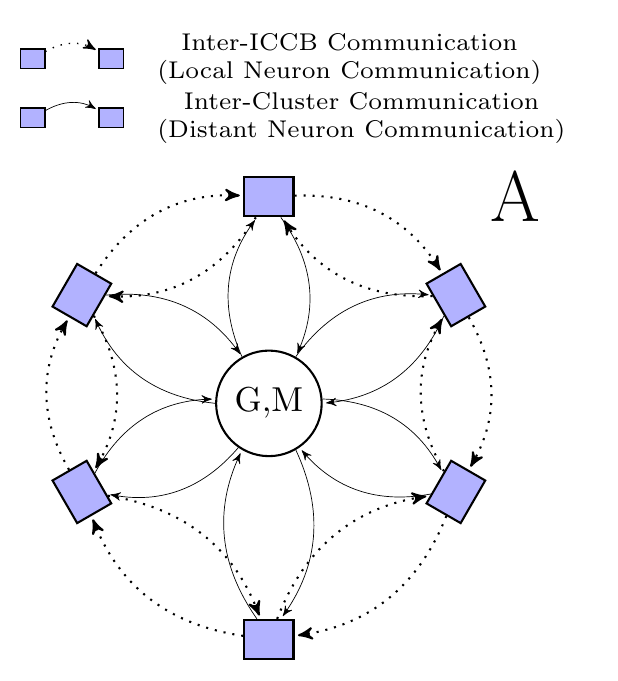}
		&
		\includegraphics[width=.50\textwidth]{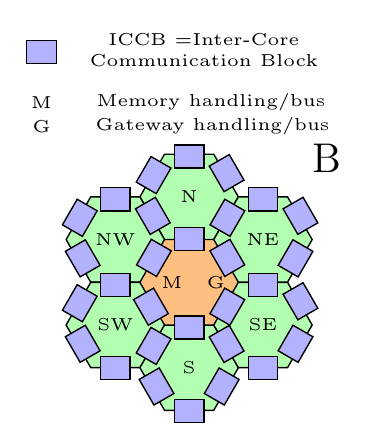}
	\end{tabular}
	\caption{
		Subfigure A (compare to Fig.7 in~\cite{BuzsakiGammaOscillations:2012}) shows a proposal~\cite{VeghSPAEMPA:2020} of how to reduce the limiting effect of the \gls{SPA}, via mimicking the communication between local neurons
		using direct-wired inter-core communication 
		and the communication between the farther neurons via using 
		the inter-cluster communication bus, in the cluster head. 
		Subfigure B suggests a possible implementation of the principle:
		the Inter-Core Communication Blocks represent a "local bus"
		(directly wired, with no contention), while the cores can communicate 
		with the cores in other clusters through the 'G' gateway as well as the 'M' (local and global) memory.
		\label{fig:AxonBus} 
	}
\end{figure*}

Of course, computing works through having time quanta:
what happens  within a clock period of the processor, happens "at the same time".
Given that the clock period of computers is in the range of nanoseconds, 
in the classic computing good approximation is that computing time is continuous.
Simulating many-neuron systems in \gls{SPA}, however,
one faces a lack of the cooperative behavior.
As the computing time in the artificial neurons is not 
proportional with the biological time they simulate,
these different time scales must be scaled to each other.

One possible way is to put a "time grid" on the processes
simulating biology: within a time slot, the artificial neurons would be free to compute, but at the
time boundary they would send the results of their calculation to each other.
This results in the neurons continuing their calculation
periodically from some concerted state.
Such a method of synchronization introduces a "biological clock period" that is million-fold longer than the clock period of the processor:
what happens in this "grid time", happens "at the same time".
Although this effect drastically reduces the achievable 
computing temporal performance~\textbf{\cite{VeghBrainAmdahl:2019}}, 
the synchronization principle is so common that also the special-purpose
neuromorphic chips~\cite{TrueNorth:2016,IntelLoihi:2018} use it as a built-in feature.
In their case the speed of neuronal functionality is hundreds of times higher than
that of the competing solutions, and the communication principles are slightly different (i.e., the non-payload/payload ratio is vastly different),
the performance-limiting effect of the 
"quantal nature of computing time" persists when used in extensive systems.

\section{The role of the workload on the computing efficiency\label{sec:Workload}}

As was very early predicted~\cite{AmdahlSingleProcessor67} and decades later
experimentally confirmed~\cite{ScalingParallel:1993},
the scaling of the parallelized computing is not linear.
Even, "\textit{there comes a point when using more processors \dots actually increases the execution time rather than reducing it}"~\cite{ScalingParallel:1993}. Where that point comes, depends on the workload. Paper~\textbf{\cite{VeghHowMany:2020}} discusses first/second order
approaches to explain the issue. The first order approach explains the experienced saturation, and the second order the predicted decrease.

\begin{figure*}
	\includegraphics[width=\textwidth]{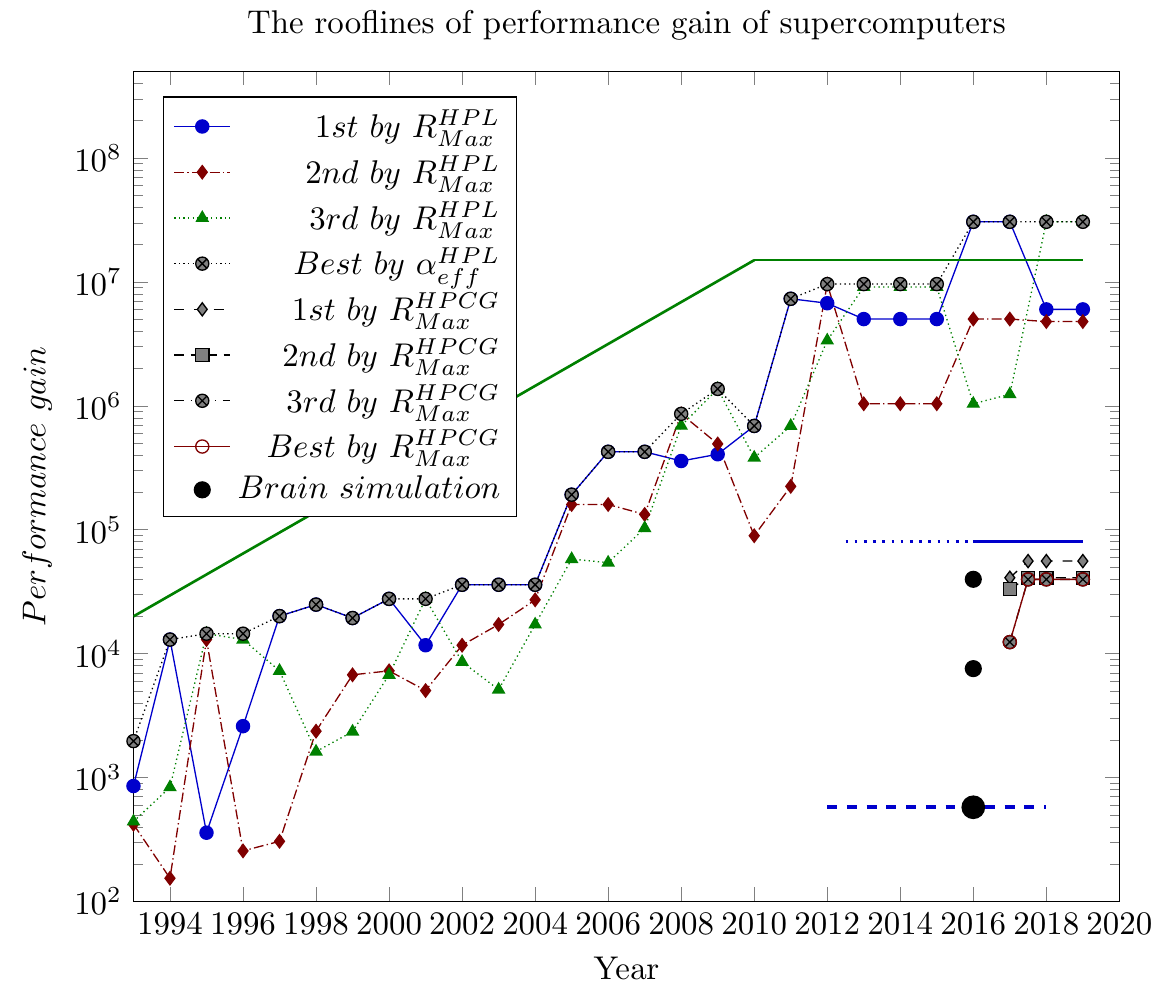}
	\caption{
	The performance gains of supercomputers modeled as "roofline"~\protect{\cite{WilliamsRoofline:2009}}, as measured with the benchmarks \gls{HPL} and \gls{HPCG} (taken from the database TOP500~\protect{\cite{TOP500:2017}}), and the one for brain simulation is concluded from
	\protect{\cite{NeuralNetworkPerformance:2018}}.
	\label{fig:PerformanceRooflines}}
\end{figure*}

As~\textbf{\cite{VeghHowMany:2020}} discusses, the different workloads, mainly due to their different communication-to-computation ratio, work with different efficiency on the same computer system~\cite{DifferentBenchmarks:2017}.
The neuromorphic operation on conventional architectures shows the same issues~\textbf{\cite{VeghAIperformance:2020,VeghReevaluate:2020,VeghScalingANN:2020}}.
Fig.~\ref{fig:PerformanceRooflines}	illustrates how the different workloads
cause a saturation in the value of the performance gain. Compared to
the benchmark \gls{HPL}, the \gls{HPCG} comprises much more communication
because of the iterative nature of the task.

Fig.~\ref{fig:PerformanceRooflines} also depicts an estimated efficiency
for the case of simulating brain-like operation on a conventional architecture.
Given that in~\cite{NeuralNetworkPerformance:2018}, the power consumption efficiency was also investigated,
one can presume that (to avoid obsolete energy consumption) the authors measured
at the point where involving more cores increased
the power consumption but did not increase the payload simulation performance.
The performance gain of an  \gls{AI} workload on supercomputers
can be estimated to be between those of \gls{HPCG} and brain simulation; closer to the \gls{HPCG} gain. As discussed experimentally in~\cite{DeepNeuralNetworkTraining:2016} and theoretically in~\textbf{\cite{VeghAIperformance:2020}}, in the case of neural networks
(especially in the case of selecting improper layering depth)
the efficiency can be much lower. Depending on the architecture,
the performance gain reaches the saturation level by using just dozens of cores in the system, mimicking neuromorphic operation on a conventional system.

\section{Limitations due to the classic computing paradigm\label{sec:ClassicComputing}}

\begin{figure*}
	\hspace{-1.1cm}\includegraphics[width=1.15\textwidth]{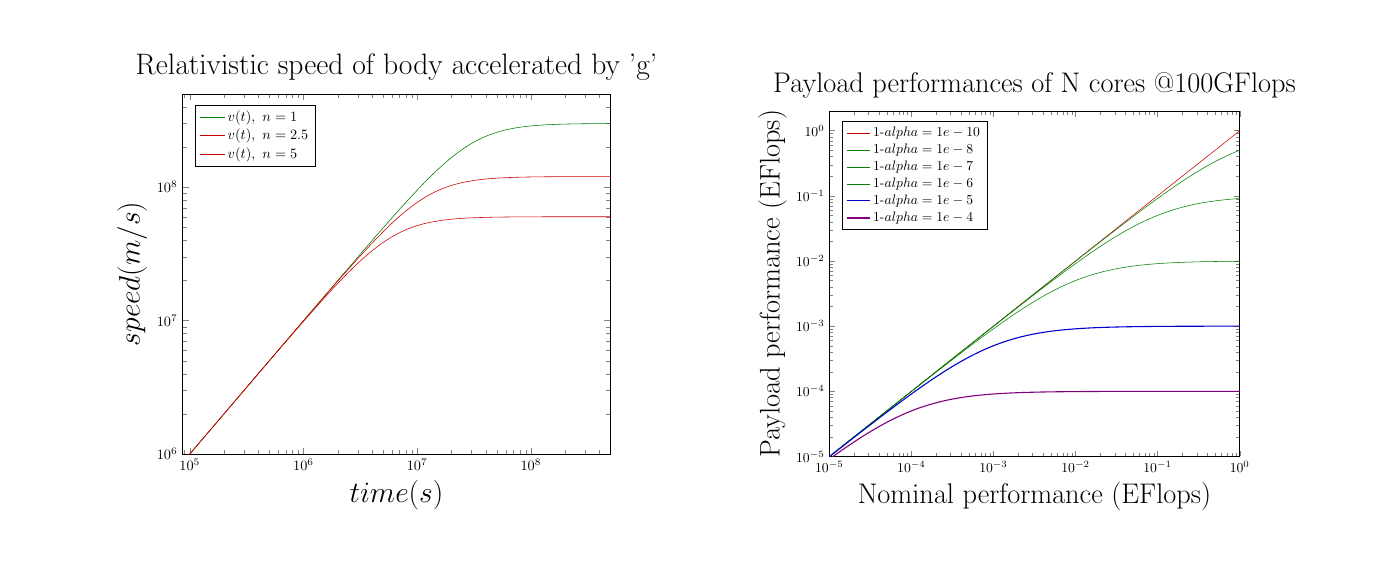}
	\vspace{-3\baselineskip}

	\caption{The limiting effect considered in the 'modern' theories.
		One left side, the speed limit, as explained by the theory of relativity, is illustrated. The refractory index of the medium defines the value of the speed limit. On the right side, the payload performance limit of the parallelized sequential computing systems, as explained by the "modern paradigm", is illustrated.
		The ratio of the non-payload to payload processing defines the payload performance.
		\label{fig:RelativisticVsPerformance}}
	\vspace{-\baselineskip}
\end{figure*}

Mainly, the effect of the shared bus defines the payload performance of the computing systems assembled from components manufactured for \gls{SPA} systems. The right subfigure in Fig.~\ref{fig:RelativisticVsPerformance} displays the payload
	performance of a many-processor \gls{SPA} system when executing different workloads (that define the non-payload to payload ratio); for the math details see~\textbf{\cite{VeghModernParadigm:2019,VeghHowMany:2020}}.
	The top diagram lines represent the best payload performance that the supercomputers can achieve when running the benchmark \gls{HPL}
	that represents the minimum communication a parallelized sequential system needs. The bottom diagram line represents the estimation of the payload performance that neuromorphic-type processing can achieve in \gls{SPA} systems.
	Notice the similarity with the left subfigure: under extreme conditions, in the science, an environment-dependent speed limit exists, and in computing, a workload-dependent payload performance limit exists~\textbf{\cite{VeghModernParadigm:2019}}.

The careful analysis discovers a remarkable parallel between the proposed 'modern computing'~\textbf{\cite{VeghModernParadigm:2019}} versus the classic computing and the modern science versus the classic science.
"Modern computing" does not invalidate the
"classic computing". Instead, it draws the range of the validity of the classic \textit{approximation} and sets up the rules of computing under extreme conditions. The \textit{"modern computing" in its field leads to counter-intuitive and shocking conclusions at some extreme parameter values}, as did the "modern science" more than hundred years ago. The parallel can help to accept that \textit{what one can not experience in the every-day computing, can be true when using computing under extreme conditions}.
Fig.~\ref{fig:RelativisticVsPerformance} depicts one such consequence.
In the modern science, unlike in the classic science, a speed limit exists.
In the modern computing, unlike in the classic computing,
a payload performance limit exists. For further parallels, see~\textbf{\cite{VeghModernParadigm:2019}}.

Using another computing theory is a must, especially when targeting neuromorphic computing.
\textit{In the frames of "classic computing"}, as was bitterly admitted~\cite{NeuralNetworkPerformance:2018}, \textit{"any
	studies on processes like plasticity, learning, and development
	exhibited over hours and days of biological time are outside our
	reach".}

\section{A system is not a simple sum of its components\label{sec:NotSimple}}

Although it is valid for most systems, that one must not conclude
from a feature of a component to the similar feature of the system:
the non-linearity discussed above it is especially valid
for the large-scale computing systems mimicking neuromorphic operation.
We mention two prominent examples here. One can assume that
if the time of the operation of a neuron can be shortened, the
performance of the whole system gets proportionally better.
Two distinct options are to use shorter operands (move less data
and to perform less bit manipulations) and to mimic the operation
of the neuron in an entirely different way: using quick analog
signal processing rather than slow digital calculation.

The so-called \textit{HPL-AI} benchmark used \textit{Mixed Precision}\footnote{Both names are used rather inconsequentially. On one side, the test itself has not much to do with \gls{AI}, just uses the operand length common in \gls{AI} tasks; the benchmark \gls{HPL}, similarly to \gls{AI}, is a workload type. On the other side, the Mixed Precision is Half Precision: it is natural that for multiplication twice as long operands are used temporarily.  It is a different question that the operations are contracted.}
\cite{MixedPrecisionHPL:2018} rather than Double Precision operands
in benchmarking their supercomputer. The name suggests as if in solving \gls{AI} tasks,
the supercomputer can show that peak efficiency.
When executing the \gls{HPL} benchmark, this change resulted in a higher performance merit number. However,
as correctly stated in the announcement,
"\textit{Achieving a 445 petaflops mixed-precision result on HPL (equivalent to our 148.6 petaflops DP result)}", i.e. the peak DP performance did not change.

We expect that when using half-precision (FP16) rather than double precision
(FP64) operands in the calculations, four times less
data are transferred and manipulated by the system.
The measured power consumption data underpin the statement.
However, the computing performance is only 3 times higher than in the case of using 64-bit (FP64) operands.
The non-linearity has its effect even in this simple case. In the benchmark, the housekeeping activity also takes time. Even
the measured performance data enable us to estimate the execution time with zero precision (FP0) operands, see~\textbf{\cite{VeghHowMany:2020}}.
The performance corresponding to $\alpha_{HPL}^{FP0}$ is slightly above 1 EFlops (when making no floating operations, i.e., rather Eops). Another peak performance reported\footnote{https://www.olcf.ornl.gov/2018/06/08/genomics-code-exceeds-exaops-on-summit-supercomputer/} when running genomics code on the same supercomputer (by using a mixture of operands with different numerical precision and mostly non-floating point instructions) is 1.88 Eops, corresponding to $\alpha_{Genom}^{FP0} = 1*10^{-8}$; for the scaling of that type of calculations see Fig.~\ref{fig:RelativisticVsPerformance}. Given that those two values refer to a different mixture of instructions,
the agreement is more than satisfactory.

Another plausible assumption is that if we use quick analog signal processing
to replace the slow digital calculation, as proposed in~\cite{RecipeMemristor:2020,NatureBuildingBrain:2020},
 the system gets proportionally quicker. Adding analog components to a digital processor, however, has its price. Given that the digital processor cannot handle resources outside of its world, one must call the \gls{OS} for help. That help, however, is rather expensive in terms of execution time. The required context switching
takes time in the order of executing $10^4$ instructions~\cite{armContextSwitching:2007,Tsafrir:2007}, which greatly increases
the total execution time and makes the non-payload to payload ratio much worse.

Although these cases seem to be very different, they share at least the common feature, that they change not only one parameter: \textit{they also change the
non-payload to payload ratio} that defines the efficiency.
They have different side-effects: changing the operand length has its effect on the cache behavior, using analog processing needs linking between the analog and the digital processing.
\textit{However, even those one-parameter changes have a nonlinear effect on the efficiency of the system.}

\section{Summary\label{sec:Summary}}

The authors have identified some critical bottlenecks in current computational systems/neuronal networks rendering the conventional computing architectures unadaptable to large (and even medium) sized neuromorphic computing. Built with the segregated processor (\gls{SPA}, wording from Amdahl~\cite{AmdahlSingleProcessor67}), the current systems lack autonomous communication of processors and have an inefficient method of imitating biological systems. To overcome these limitations, the authors introduce a \textit{drastically different approach to computing}, the Explicitly Many-Processor Approach (EMPA), which can serve as the basis for development as well as specific and practical solutions.

\section*{Acknowledgemens}
The authors thank Prof. P\'eter Somogyi for valuable comments on a previous version of the manuscript. 
Project no. 125547  should have been implemented with the support provided from the National Research, Development and Innovation Fund of Hungary, financed under the K funding scheme. However, at the time of writing the paper,
the fund is in 20 month delay with providing the support.
Because of this, temporarily, the project is supported by the Kalimános BT.
%Also the ERC-ECAS support of project 886183 is acknowledged.
%Project no. 125547  has been implemented with the support provided from the National Research, Development and Innovation Fund of Hungary, financed under the K funding scheme.


\begin{thebibliography}{}
\expandafter\ifx\csname url\endcsname\relax
  \def\url#1{\texttt{#1}}\fi
\expandafter\ifx\csname urlprefix\endcsname\relax\def\urlprefix{URL }\fi
\expandafter\ifx\csname href\endcsname\relax
  \def\href#1#2{#2} \def\path#1{#1}\fi

\end{thebibliography}


\section*{References}

\begin{thebibliography}{10}
	\expandafter\ifx\csname url\endcsname\relax
	\def\url#1{\texttt{#1}}\fi
	\expandafter\ifx\csname urlprefix\endcsname\relax\def\urlprefix{URL }\fi
	\expandafter\ifx\csname href\endcsname\relax
	\def\href#1#2{#2} \def\path#1{#1}\fi
	
	\bibitem{NeuromorphicSystems:1990}
	C.~Mead, {Neuromorphic electronic systems}, {Proc. IEEE} 78 (1990) 1629--1636.
	
	\bibitem{NeuromorphicComputing:2015}
	{US DOE Office of Science}, {Report of a Roundtable Convened to Consider
		Neuromorphic Computing Basic Research Needs},
	\url{https://science.osti.gov/-/media/ascr/pdf/programdocuments/docs/Neuromorphic-Computing-Report\_FNLBLP.pdf}
	(2015).
	
	\bibitem{GordonBellPrize:2017}
	G.~Bell, D.~H. Bailey, J.~Dongarra, A.~H. Karp, K.~Walsh,
	\href{https://doi.org/10.1177/1094342017738610}{{A look back on 30 years of
			the Gordon Bell Prize}}, The International Journal of High Performance
	Computing Applications 31~(6) (2017) 469–484.
	\newline\urlprefix\url{https://doi.org/10.1177/1094342017738610}
	
	\bibitem{FurberNeuralEngineering:2007}
	{Steve Furber and Steve Temple}, {Neural systems engineering}, {J. R. Soc.
		Interface} 4 (2007) 193--206.
	\newblock \href {http://dx.doi.org/10.1098/rsif.2006.0177}
	{\path{doi:10.1098/rsif.2006.0177}}.
	
	\bibitem{SpiNNaker2:2018}
	C.~Liu, G.~Bellec, B.~Vogginger, D.~Kappel, J.~Partzsch, F.~Neumärker,
	S.~Höppner, W.~Maass, S.~B. Furber, R.~Legenstein, C.~G. Mayr,
	\href{https://www.frontiersin.org/article/10.3389/fnins.2018.00840}{{Memory-Efficient
			Deep Learning on a SpiNNaker 2 Prototype}}, Frontiers in Neuroscience 12
	(2018) 840.
	\newblock \href {http://dx.doi.org/10.3389/fnins.2018.00840}
	{\path{doi:10.3389/fnins.2018.00840}}.
	\newline\urlprefix\url{https://www.frontiersin.org/article/10.3389/fnins.2018.00840}
	
	\bibitem{DOEAurora:2017}
	{Top500.org}, {Retooled Aurora Supercomputer Will Be America’s First Exascale
		System},
	\url{https://www.top500.org/news/retooled-aurora-supercomputer-will-be
		-americas- first-exascale-system/} (2017).
	
	\bibitem{VeghReevaluate:2020}
	J.~V\'egh, \href{https://arxiv.org/abs/2002.08316}{{Re-evaluating scaling
			methods for distributed parallel systems}}, IEEE Transactions on Distributed
	and Parallel Computing ?? (2020) in review.
	\newline\urlprefix\url{https://arxiv.org/abs/2002.08316}
	
	\bibitem{VeghScalingANN:2020}
	J.~V\'egh, {Which scaling rule applies to Artificial Neural Networks}, in:
	{2020 International Conference on Computational Science and Computational
		Intelligence (CSCI)}, IEEE, 2020, p. Submitted.
	
	\bibitem{DeepNeuralNetworkTraining:2016}
	J.~Keuper, F.-J. Preundt,
	\href{https://www.researchgate.net/publication/308457837}{{Distributed
			Training of Deep Neural Networks: Theoretical and Practical Limits of
			Parallel Scalability}}, in: 2nd Workshop on Machine Learning in HPC
	Environments (MLHPC), IEEE, 2016, pp. 1469--1476.
	\newblock \href {http://dx.doi.org/10.1109/MLHPC.2016.006}
	{\path{doi:10.1109/MLHPC.2016.006}}.
	\newline\urlprefix\url{https://www.researchgate.net/publication/308457837}
	
	\bibitem{VeghAIperformance:2020}
	J.~V\'egh, {How deep the machine learning can be}, A Closer Look at
	Convolutional Neural Networks, Nova, In press, 2020, pp. 141--169.
	
	\bibitem{AsanovicParallelCACM:2009}
	K.~Asanovic, R.~Bodik, J.~Demmel, T.~Keaveny, K.~Keutzer, J.~Kubiatowicz,
	N.~Morgan, D.~Patterson, K.~Sen, J.~Wawrzynek, D.~Wessel, K.~Yelick, {A View
		of the Parallel Computing Landscape}, Comm. ACM 52~(10) (2009) 56--67.
	
	\bibitem{SoOS:2010}
	{S(o)OS~project}, Resource-independent execution support on exa-scale systems,
	\url{http://www.soos-project.eu/index.php/related-initiatives} (2010).
	
	\bibitem{DeBenedictis_supercomputing:2014}
	{Machine Intelligence Research Institute},
	\href{https://intelligence.org/2014/04/03/erik-debenedictis/}{{Erik
			DeBenedictis on supercomputing}} (2014).
	\newline\urlprefix\url{https://intelligence.org/2014/04/03/erik-debenedictis/}
	
	\bibitem{RebootingComputingModels:2019}
	P.~C. et~al., {Rebooting Our Computing Models}, in: Proceedings of the 2019
	Design, Automation \& Test in Europe Conference \& Exhibition (DATE), IEEE
	Press, 2019, pp. 1469--1476.
	\newblock \href {http://dx.doi.org/10.23919/DATE.2019.8715167}
	{\path{doi:10.23919/DATE.2019.8715167}}.
	
	\bibitem{WilliamsRoofline:2009}
	S.~Williams, A.~Waterman, D.~Patterson, Roofline: An insightful visual
	performance model for multicore architectures, Commun. ACM 52~(4) (2009)
	65--76.
	
	\bibitem{Gustafson:1988}
	J.~L. Gustafson, {Reevaluating Amdahl's Law}, Commun. ACM 31~(5) (1988)
	532--533.
	\newblock \href {http://dx.doi.org/10.1145/42411.42415}
	{\path{doi:10.1145/42411.42415}}.
	
	\bibitem{AmdalVsGustafson96}
	Y.~Shi, {Reevaluating Amdahl's Law and Gustafson's Law},
	\url{https://www.researchgate.net/publication/
		228367369\_Reevaluating\_Amdahl's\_law\_and\_Gustafson's\_law} (1996).
	
	\bibitem{PerformanceCounter2013}
	V.~Weaver, D.~Terpstra, S.~Moore, Non-determinism and overcount on modern
	hardware performance counter implementations, in: Performance Analysis of
	Systems and Software (ISPASS), 2013 IEEE International Symposium on, 2013,
	pp. 215--224.
	\newblock \href {http://dx.doi.org/10.1109/ISPASS.2013.6557172}
	{\path{doi:10.1109/ISPASS.2013.6557172}}.
	
	\bibitem{VeghHowMany:2020}
	J.~V{\'{e}}gh, Finally, how many efficiencies the supercomputers have?, The
	Journal of Supercomputing\href {http://dx.doi.org/10.1007/s11227-020-03210-4}
	{\path{doi:10.1007/s11227-020-03210-4}}.
	
	\bibitem{ScalingParallel:1993}
	J.~P. Singh, J.~L. Hennessy, A.~Gupta, Scaling parallel programs for
	multiprocessors: Methodology and examples, Computer 26~(7) (1993) 42--50.
	\newblock \href {http://dx.doi.org/10.1109/MC.1993.274941}
	{\path{doi:10.1109/MC.1993.274941}}.
	
	\bibitem{BuzsakiGammaOscillations:2012}
	{Gy\"orgy Buzs\'aki and Xiao-Jing Wang}, {Mechanisms ofGamma Oscillations},
	{Annual Reviews of Neurosciences} 3~(4) (2012) 19:1--19:29.
	\newblock \href {http://dx.doi.org/10.1146/annurev-neuro-062111-150444}
	{\path{doi:10.1146/annurev-neuro-062111-150444}}.
	
	\bibitem{VeghSPAEMPA:2020}
	J.~V\'egh, \href{??http://arxiv.org/abs/1908.02651}{{How to extend the
			Single-Processor Paradigm to the Explicitly Many-Processor Approach}}, in:
	{2020 International Conference on Computational Science and Computational
		Intelligence (CSCI)}, IEEE, 2020, p. In print.
	\newline\urlprefix\url{??http://arxiv.org/abs/1908.02651}
	
	\bibitem{VeghBrainAmdahl:2019}
	{J. V\'egh},
	\href{https://braininformatics.springeropen.com/articles/10.1186/s40708-019-0097-2/metrics}{{How
			Amdahl's Law limits the performance of large artificial neural networks:
			{\small (Why the functionality of full-scale brain simulation on
				processor-based simulators is limited)}}}, Brain Informatics 6 (2019) 1--11.
	\newline\urlprefix\url{https://braininformatics.springeropen.com/articles/10.1186/s40708-019-0097-2/metrics}
	
	\bibitem{TrueNorth:2016}
	J.~S. et~al, {TrueNorth Ecosystem for Brain-Inspired Computing: Scalable
		Systems, Software, and Applications}, in: SC '16: Proceedings of the
	International Conference for High Performance Computing, Networking, Storage
	and Analysis, 2016, pp. 130--141.
	
	\bibitem{IntelLoihi:2018}
	M.~{Davies, et al}, {Loihi: {\small A Neuromorphic Manycore Processor with
			On-Chip Learning}}, IEEE Micro 38 (2018) 82--99.
	
	\bibitem{AmdahlSingleProcessor67}
	G.~M. Amdahl, {Validity of the Single Processor Approach to Achieving
		Large-Scale Computing Capabilities}, in: AFIPS Conference Proceedings,
	Vol.~30, 1967, pp. 483--485.
	\newblock \href {http://dx.doi.org/10.1145/1465482.1465560}
	{\path{doi:10.1145/1465482.1465560}}.
	
	\bibitem{TOP500:2017}
	TOP500, {November 2017 list of supercomputers},
	\url{https://www.top500.org/lists/2017/11/} (2017).
	
	\bibitem{NeuralNetworkPerformance:2018}
	S.~J. van Albada, A.~G. Rowley, J.~Senk, M.~Hopkins, M.~Schmidt, A.~B. Stokes,
	D.~R. Lester, M.~Diesmann, S.~B. Furber, {Performance Comparison of the
		Digital Neuromorphic Hardware SpiNNaker and the Neural Network Simulation
		Software NEST for a Full-Scale Cortical Microcircuit Model}, Frontiers in
	Neuroscience 12 (2018) 291.
	
	\bibitem{DifferentBenchmarks:2017}
	{IEEE Spectrum}, {Two Different Top500 Supercomputing Benchmarks Show Two
		Different Top Supercomputers},
	\url{https://spectrum.ieee.org/tech-talk/computing/hardware/two-different-top500-supercomputing-
		benchmarks-show\ -two -different-top-supercomputers} (2017).
	
	\bibitem{VeghModernParadigm:2019}
	J.~{Végh}, A.~{Tisan}, \href{http://arxiv.org/abs/1908.02651}{{The need for
			modern computing paradigm: Science applied to computing}}, in: 2019
	International Conference on Computational Science and Computational
	Intelligence (CSCI), IEEE, 2019, pp. 1523--1532.
	\newblock \href {http://dx.doi.org/10.1109/CSCI49370.2019.00283}
	{\path{doi:10.1109/CSCI49370.2019.00283}}.
	\newline\urlprefix\url{http://arxiv.org/abs/1908.02651}
	
	\bibitem{MixedPrecisionHPL:2018}
	A.~Haidar, S.~Tomov, J.~Dongarra, N.~J. Higham, {Harnessing GPU Tensor Cores
		for Fast FP16 Arithmetic to Speed Up Mixed-precision Iterative Refinement
		Solvers}, in: Proceedings of the International Conference for High
	Performance Computing, Networking, Storage, and Analysis, SC '18, IEEE Press,
	2018, pp. 47:1--47:11.
	
	\bibitem{RecipeMemristor:2020}
	E.~Chicca, G.~Indiveri, \href{https://doi.org/10.1063/1.5142089}{{A recipe for
			creating ideal hybrid memristive-CMOS neuromorphic processing systems}},
	Applied Physics Letters 116~(12) (2020) 120501.
	\newblock \href {http://arxiv.org/abs/https://doi.org/10.1063/1.5142089}
	{\path{arXiv:https://doi.org/10.1063/1.5142089}}, \href
	{http://dx.doi.org/10.1063/1.5142089} {\path{doi:10.1063/1.5142089}}.
	\newline\urlprefix\url{https://doi.org/10.1063/1.5142089}
	
	\bibitem{NatureBuildingBrain:2020}
	\href{https://doi.org/10.1038/s41467-019-12521-x}{{Building brain-inspired
			computing}}, Nature Communications 10~(12) (2019) 4838.
	\newblock \href {http://dx.doi.org/10.1063/1.5142089}
	{\path{doi:10.1063/1.5142089}}.
	\newline\urlprefix\url{https://doi.org/10.1038/s41467-019-12521-x}
	
	\bibitem{armContextSwitching:2007}
	F.~M. David, J.~C. Carlyle, R.~H. Campbell,
	\href{http://doi.acm.org/10.1145/1281700.1281703}{{Context Switch Overheads
			for Linux on ARM Platforms}}, in: Proceedings of the 2007 Workshop on
	Experimental Computer Science, ExpCS '07, ACM, New York, NY, USA, 2007.
	\newblock \href {http://dx.doi.org/10.1145/1281700.1281703}
	{\path{doi:10.1145/1281700.1281703}}.
	\newline\urlprefix\url{http://doi.acm.org/10.1145/1281700.1281703}
	
	\bibitem{Tsafrir:2007}
	D.~Tsafrir, The context-switch overhead inflicted by hardware interrupts (and
	the enigma of do-nothing loops), in: Proceedings of the 2007 Workshop on
	Experimental Computer Science, ExpCS '07, ACM, New York, NY, USA, 2007, pp.
	3--3.
	
\end{thebibliography}
\end{document}